\documentclass[aps,prd,showpacs,amsmath,amssymb,nofootinbib,twocolumn]{revtex4-1}
\usepackage{graphicx}
\usepackage{graphicx,array,dcolumn}
\usepackage{calc,tabularx, epsfig,mathrsfs}
\usepackage{hyperref}
\usepackage{natbib}

\usepackage{amsmath,verbatim,enumerate}
\usepackage{amssymb}
\usepackage{multirow}

\usepackage{bm}

\def\beq{\begin{equation}}
\def\eeq{\end{equation}}

\def\bal{\begin{align}}

\def\eal{\end{align}}
\begin{document}

\title{Bulk-boundary dictionary for a non local bulk field}

\author{Nirmalya Kajuri}
\email{nkajuri@cmi.ac.in}
\affiliation{Chennai Mathematical Institute, Siruseri
Kelambakkam 603103}
\author{Gaurav Narain}
\email{gnarain@buaa.edu.cn}
\affiliation{Department of Space Science, Beihang University, Beijing 100191, China.}

\bibliographystyle{apsrev4-1}

\begin{abstract}

AdS/CFT correspondence gives us a bulk-boundary dictionary between CFT operators and local fields in the bulk. Can a bulk-boundary dictionary exist for a non local bulk field? In this paper we consider a particular non local theory and show that a bulk boundary dictionary relating it to CFT operators in the bulk can indeed be constructed. In this dictionary the scale of non locality of the bulk theory is related to the conformal dimension of the dual CFT operator.

\end{abstract}

\maketitle

\section{Introduction.} 
\label{intro}

According to the AdS/CFT correspondence\cite{Maldacena:1997re,Gubser:1998bc, Witten:1998qj}, there is a duality between local fields in the bulk and the operators in the boundary CFT. For a scalar field in a $d+1$ dimensional bulk this duality
can be stated in the form of the following extrapolate dictionary\cite{Banks:1998dd,Balasubramanian:1998sn}:

\begin{align}
\label{xp}\notag\lim_{r\to\infty}r^{n\Delta}\langle\phi(r_1,x_1)\phi(r_2,x_2).....\phi(r_n,x_n)\rangle=\\ \langle 0|\mathcal{O}(x_1)\mathcal{O}(x_2)....\mathcal{O}(x_n)|0\rangle
\end{align}

Here $\mathcal{O}$ is a primary operator in the boundary CFT whose dimension $\Delta$ is related to the mass of the bulk scalar field  
\beq \label{stand}
\Delta = \frac{d}{2} +\frac{1}{2}\sqrt{d^2+4m^2} 
\eeq
A similar dictionary holds for other fields.

It is interesting to ask about the possibility of boundary duals to non local fields in the bulk. Can a bulk-boundary duality like \eqref{xp} exist for a non local bulk field? If not, why exactly would the correlators of a non local bulk field fail to satisfy \eqref{xp}? If boundary duals to non local fields do exist, can one tell from the boundary theory if its bulk dual is local or not?

One motivation to ask these questions is to clarify the understanding of bulk locality in AdS/CFT. It is not yet fully understood how locality of bulk field theories is encoded in the dual CFT \cite{Heemskerk:2009pn, ElShowk:2011ag}.

Investigating CFT duals for non local bulk theories could help shed light on the issue of bulk locality. For instance, a non local theory would have a scale of non locality. If a bulk-boundary dictionary can be constructed for non local fields, it would tell us what the scale of non locality translates to in the boundary. 

A second motivation to consider boundary duals for non local bulk fields comes from the work of Akhmedov et al \cite{Akhmedov:2018lkp} who have shown that non local counterterms arise naturally when one considers one loop corrections to the four-point function in a $\phi^4$ theory on AdS background. This work suggests that the question of boundary duals for non local bulk fields has to be explored further.
 
Non local terms also appear naturally in effective actions for perturbative quantum gravity at higher orders (See for instance \cite{Barvinsky:1985an, Buchbinder:1992rb,Barvinsky:1993en,Donoghue:2014yha, Calmet:2015dpa} and \cite{Donoghue:2017pgk} for a review). Finally, non local theories have recently seen a resurgence in various contexts \cite{Moffat:1990jj,Tomboulis:1997gg,Nojiri:2007uq, Deser:2007jk,Barnaby:2007ve,Barnaby:2008fk,Jhingan:2008ym,Capozziello:2008gu,Elizalde:2011su,Biswas:2010zk,Moffat:2010bh,Barnaby:2010kx, Calcagni:2010ab,Nojiri:2010wj,Biswas:2011ar,Modesto:2011kw,Elizalde:2011su, Deser:2013uya,Aslanbeigi:2014zva, Belenchia:2014fda, Modesto:2017sdr,Maggiore:2016gpx,Kajuri:2017jmy,Narain:2017twx,Calcagni:2017sov,Calcagni:2018gke, Calcagni:2018lyd,Saravani:2018rwm, Narain:2018hxw, Kajuri:2018myh}. Finding a boundary dual can help improve our understanding of non local field theories themselves.

In this paper we take the first steps towards studying non local theories in a holographic context. We consider a particular non local field theory and we show that a bulk-boundary dictionary of the form \eqref{xp} can be constructed for this theory. However it is not the exact same dictionary in that the conformal dimension of the dual operator has a different relation with the scales of the bulk theory. The conformal dimension $\Delta$ is found to be related to both the mass and the scale of non locality of the bulk theory. Other than that, the bulk-boundary dictionary is identical. We call it the modified extrapolate dictionary. The existence of this dictionary is our main result.

We arrive at the modified extrapolate dictionary by starting with the two-point function of the free non local field $\phi$ and asking if there is a $\Delta$ such that in the boundary limit $r^{\Delta}\langle \phi(r_1,x_1) \phi(r_2,x_2) \rangle$ gives a conformal two-point function. We then introduce interactions and check that the same prescription holds for higher-point functions.

The paper is organized as follows. In the next section we introduce a free non local field theory in question. In section III we will find its CFT dual. We will include interactions in section IV. The final section presents our conclusions.

\section{ A non local field theory}
\label{simpNL}

We start with a free massless non local theory on $AdS_{d+1}$ space-time with the following action
\beq \label{nlaction}
S = \int d^{d+1}x \sqrt{-g} 
\left[\frac{1}{2}(\partial_\mu \phi)^2  - \frac{\lambda^2}{2} \phi \frac{1}{\Box} \phi \right] \, .
\eeq
Here $\frac{1}{\Box}$ denotes the D'Alembartian in AdS space, $\phi$ is the bulk scalar field while $g=\det(g_{\mu\nu})$. The parameter $\lambda$ dictates the strength of non-locality of the theory and has mass dimensions $M^2$. It will henceforth be referred to as the non locality scale.

This theory can be re-written as a local action by introducing an extra field:
\beq \label{laction}
S = \int d^{d+1}x  \sqrt{-g}\left[ \frac{1}{2}(\partial_\mu \phi)^2 2+ \frac{1}{2}(\partial_\mu \chi)^2 + \lambda \phi \chi \right] \, .
\eeq
Integrating out $\chi$ gives us the non local theory of \eqref{nlaction}. 

What makes this action particularly interesting is that a further field redefinition turns it into two uncoupled local theories. By making the following field redefinitions

\beq
\label{frdef}
\notag \phi = \frac{1}{\sqrt{2}}(\psi_1 +\psi_2) \, ,
\hspace{5mm}
\chi = \frac{1}{\sqrt{2}}(\psi_1 -\psi_2) \, .
\eeq

and substituting these in \eqref{laction} we get a theory of two uncoupled scalars:
\beq  
\begin{aligned}
\label{uncoup}
\notag S = \int d^{d+1}x \sqrt{-g} \left[
  \left((\partial_\mu \psi_1)^2 +  \lambda \psi_1^2\right)\right. \\+\left. B\left((\partial_\mu \psi_2)^2 -\lambda\psi_2^2 \right)\right]\, .
\end{aligned}
\eeq

So we find that $\psi_2$ is a tachyonic field. This however is not problematic as long as the BF bound is satisfied 
\beq 
\label{bf}
\lambda  < \frac{d^2}{4}
\eeq 
So for our non local theory to be sensible in AdS (and have a sensible CFT dual) it has to satisfy a restriction on the scale of non locality $\lambda$. In what follows we will assume that \eqref{bf} holds.

We can also see from the above that the non local theory \eqref{nlaction} is both causal and unitary, because the local theory \eqref{uncoup} is.

\section{CFT dual to the free non local theory}

We now investigate the possibility of constructing a CFT dual for these theories.

The two point function of the field $\phi$ can be written in terms of these new fields:

\begin{align}
\label{phipsi}
\notag & 2\langle\phi(r_1,x_1)\phi(r_2,x_2)\rangle= \\&\frac{1}{2}\langle\psi_1(r_1,x_1)\psi_1(r_2,x_2)\rangle +\frac{1}{2}\langle\psi_2(r_1,x_1)\psi_2(r_2,x_2)\rangle
\end{align}

Now the terms on the right hand side are correlators of free local scalar fields on AdS. We know how to relate them to CFT correlators in the boundary limit from \eqref{xp}. These are given by:
\beq
\label{eq:rel1}
\lim_{r \to \infty} \langle\psi_1(r_1,x_1)\psi_1(r_2,x_2)\rangle \approx  r^{-2\Delta_1} \langle\mathcal{O}_{\Delta_1}(x_1)\mathcal{O}_{\Delta_1}(x_2)\rangle
\eeq
and
\beq
\label{eq:rel2}
\lim_{r \to \infty}  \langle\psi_2(r_1,x_1)\psi_2(r_2,x_2)\rangle \approx  r^{-2\Delta_2} \langle\mathcal{O}_{\Delta_2}(x_1)\mathcal{O}_{\Delta_2}(x_2)\rangle \, ,
\eeq
We have used the notation $\mathcal{O}_{\Delta}$ to denote a primary operator of conformal dimension $\Delta$ and 
\beq
\label{nlconf1}
\Delta_1 =  \frac{d}{2} +\frac{1}{2}\sqrt{d^2+4\lambda}
\eeq

\beq 
\label{nlconf2}
\Delta_2 =  \frac{d}{2} +\frac{1}{2}\sqrt{d^2- 4\lambda}
\eeq

Substituting this in \eqref{phipsi} gives us the large distance behaviour of the $\langle \phi \phi \rangle$ correlator\footnote{In this paper we have only considered the $r \to \infty$ limit of the two-point function of the theory \eqref{nlaction}. The general expression for the two-point function is given in \cite{Narain:2018nfh}}:
\begin{align}
\label{eq:LDphiphi}
\notag &\langle\phi(r_1,x_1)\phi(r_2,x_2)\rangle =\\ & A r^{-2\Delta_1}  \langle\mathcal{O}_{\Delta_1}(x_1)\mathcal{O}_{\Delta_1}(x_2)\rangle + B r^{-2\Delta_2}  \langle\mathcal{O}_{\Delta_2}(x_1)\mathcal{O}_{\Delta_2}(x_2)\rangle
\end{align}

Is it possible to define a CFT dual to this theory? Note that to get an equation like \eqref{xp} we have to multiply the LHS by a suitable factor and then take $r\to\infty$ limit such that the LHS is a CFT correlator. From the above equation, we can see that there is a unique way of doing this. We must multiply the LHS by $r^{2\Delta_2}$ and take the $r\to\infty$ limit. Then we indeed get a CFT correlator on the RHS :

\beq 
\label{result}
r^{2\Delta_2}\langle\phi(r_1,x_1)\phi(r_2,x_2)\rangle =  \langle\mathcal{O}_{\Delta_2}(x_1)\mathcal{O}_{\Delta_2}(x_2)\rangle
\eeq

Here we absorbed the factor of $\frac{1}{2}$ in the normalization of the CFT two-point function. Thus we found that it is indeed possible to define an extrapolate dictionary for the non local theory \eqref{nlaction}, which has the same form as the usual extrapolate dictionary except. The difference is that the conformal dimension of the dual primary is now related to the non locality scale $\lambda$ through \eqref{nlconf2}.

We call the relation \eqref{result} (and its generalization to higher point functions) as the modified extrapolate dictionary.

\section{Including Interactions}

In the last section we found the scaling prescription \eqref{result} for constructing CFT correlators from the correlators of the free non local field $\phi$. However, it could well be that this holds only at the free field level and fails when interactions are introduced. We would now like introduce prescriptions and check if the dictionary holds in the presence of interactions.

Let us introduce a general monomial interaction term :

\beq 
\mathcal{L}_{int} = \phi^n.
\eeq
In terms of the local fields $\psi_1,\psi_2$, it is $(\frac{1}{2}\psi_1 +\frac{1}{2}\psi_2)^n$.
If the modified extrapolate dictionary holds for interacting theories, the n-point correlator of the  non local field should satisfy:

\begin{align} \label{interact}
\notag r^{n\Delta_2}\langle \phi(r_1,x_1) \phi(r_2,x_2)....\phi(r_n,x_n)\rangle = \\  \langle \mathcal{O}_{\delta_2}(x_1) \mathcal{O}_{\delta_2}(x_2).....\mathcal{O}_{\delta_2}(x_2)\rangle
\end{align}

It is easy to check that this is indeed true. We rewrite the RHS as:
\begin{align}
\notag r^{n\Delta_2}\langle (\frac{1}{2}\psi_1(r_1,x_1) +\frac{1}{2}\psi_2(r_1,x_1)) (\frac{1}{2}\psi_1(r_2,x_2) \\ +\frac{1}{2}\psi_2(r_2,x_2))....(\frac{1}{2}\psi_1(r_n,x_n) +\frac{1}{2}\psi_2(r_n,x_n))\rangle
\end{align} 

Then we expand this out as a sum of correlators and use the extrapolate dictionary for each correlator. It is easy to see that the only term that does not go to zero in the boundary limit is the one which falls off as $r^{-n\delta_2}$. This is precisely the term in the right hand side of \eqref{interact} (the factors of $\frac{1}{2}$ get absorbed because of our choice of normalization. Clearly, this would be go through for general polynomial interaction terms.

We can also introduce different non local fields $\phi_A, \phi_B,...$ and consider interactions of the form $\phi_A\phi_B..$. One can follow the same steps as above and check that the non local extrapolate dictionary holds for this case as well.

Thus we find that the modified extrapolate dictionary holds in general for the non local field $\phi$:

\begin{align} \label{final}
\notag r^{n\delta_{NL}}\langle \phi(r_1,x_1) \phi(r_2,x_2)....\phi(r_n,x_n)\rangle = \\  (-1)^n \langle \mathcal{O}_{\delta_{NL}}(x_1) \mathcal{O}_{\delta_{NL}}(x_2).....\mathcal{O}_{\delta_{NL}}(x_2)\rangle
\end{align}

where 
\beq \label{delta}
\Delta_{NL} =  \frac{d}{2} +\frac{1}{2}\sqrt{d^2- 4\lambda}
\eeq

\section{Discussion} 
\label{disc}
Our main finding is that even for a non local field theory one can construct a bulk-boundary dictionary. For the non local field theory \eqref{nlaction} we found the dictionary given by \eqref{final}. 

The modified extrapolate dictionary \eqref{final} that we found for the non local theory has one point of difference from the standard extrapolate dictionary for local fields. For local fields, the conformal dimension of the dual CFT operator is related to the mass of the field. But for our non local field theory, the conformal dimension of the dual primary (given by \eqref{delta}) is related to  the non locality scale. 

The standard extrapolate dictionary is equivalent to the statement of equality between the CFT partition function and the bulk action (in the semiclassical regime) \cite{Harlow:2011ke}. This won't be true of the non local field as because \eqref{stand} no longer holds true. It would be interesting to check if some modified version of that statement hold for the modified extrapolate dictionary.

We found that the non-locality scale is seen to be related to the conformal dimension of the dual primary in the boundary. This hints towards a boundary interpretation of non locality in terms of the conformal dimension of the dual primary. Presence of non locality lowers the conformal dimension, the same as a tachyonic mass would.

Also, the normalization of boundary primaries will be modified in the $\lambda \to 0$ limit. This is because in this limit for say the two-point function, both the two-point functions of $\psi_1$ and $\psi_2$ fall off at equal rates and therefore both contribute to the boundary two-point function. So the boundary two-point function given by the modified dictionary will be off by a factor of 2 ( and by $2^n$ for a n-point function). In this respect non locality is different from tachyonic mass and the $\lambda \to 0$ limit is not a smooth one.

These results are indicative and provide a small first step towards understanding non locality of a field theory from the boundary. Although our results only hold for a particular theory, they indicate that the subject deserves further investigation.

\begin{acknowledgements} 
NK would like to thank Suryanarayana Nemani for helpful comments.
NK is supported by the SERB National Postdoctoral Fellowship. GN is supported by grant `1000 Talent' (``Zhuoyue" Fellowship). 
\end{acknowledgements} 

\bibliography{nlads}
\end{document}